\documentclass{emulateapj}
\usepackage{bm}
\usepackage{color}
\usepackage{multirow}
\usepackage{graphicx}
\usepackage{longtable}
\usepackage{rotating}
\usepackage{threeparttable}
\usepackage{epsfig}
\shorttitle{An apparent redshift dependence of quasar continuum}
\shortauthors{Xie et al.}

\begin{document}

\title{AN APPARENT REDSHIFT DEPENDENCE OF QUASAR CONTINUUM:
IMPLICATION FOR COSMIC DUST EXTINCTION?
}

\author{Xiaoyi Xie\altaffilmark{1,2},
Shiyin Shen\altaffilmark{1,3\dag},
Zhengyi Shao\altaffilmark{1,3},
Jun Yin\altaffilmark{1},
}

\affil{\altaffilmark{1} Key Laboratory for Research in Galaxies and Cosmology,
  Shanghai Astronomical Observatory, Chinese Academy of Sciences, 80 Nandan Road, Shanghai 200030, China; ssy@shao.ac.cn}
\affil{\altaffilmark{2} Graduate University of the Chinese Academy of Sciences, No.19A Yuquan Road, Beijing 100049, China}
\affil{\altaffilmark{3} Key Lab for Astrophysics, Shanghai 200234, China}
\altaffiltext{\dag}{ssy@shao.ac.cn}

\begin{abstract}
We investigate the luminosity and redshift dependence of the quasar continuum by means of the composite spectrum using a large non-BAL radio-quiet quasar sample drawn from the Sloan Digital Sky Survey. 
Quasar continuum slopes in the UV-Opt band are measured at two different wavelength ranges, i.e., $\alpha_{\nu12}$ ($1000\sim 2000 \rm\AA$) and $\alpha_{\nu24}$ ($2000 \sim 4000 \rm\AA$) derived from a power-law fitting. 
Generally, the UV spectra slope becomes harder (higher $\alpha_{\nu}$) toward higher bolometric luminosity.
On the other hand, when quasars are further grouped into luminosity bins, we find that both $\alpha_{\nu12}$ and $\alpha_{\nu24}$ show significant anti-correlations with redshift (i.e., the quasar continuum becomes redder toward higher redshift). 
We suggest that the cosmic dust extinction is very likely the cause of this observed $\alpha_\nu-z$ relation.
We build a simple cosmic dust extinction model to quantify the observed reddening tendency and find an effective dust density $n\sigma_v \sim 10^{-5}h~\rm Mpc^{-1}$ at $z<1.5$. 
The other possibilities that could produce such a reddening effect have also been discussed.

\end{abstract}


\section{Introduction}
Observationally, quasars show very similar UV-optical spectra \citep{Francis et al.(1992)}, which can be characterized by a featureless continuum and a series of broad emission lines \citep{Peterson(1997)}.
At wavelengths longer than the Ly$\alpha$ line, from $\sim1300$ to $5000\rm\AA$, the continuum of quasars can be well described by a power-law $f_{\nu}\propto\nu^{\alpha_\nu}$,
where $\alpha_\nu$ characterizes the continuum slope with a typical value $\sim -0.5$ \citep{Vanden Berk et al.(2001)}.
In terms of ${\alpha_\nu}$, more negative ${\alpha_\nu}$ means a steeper (or redder, softer) continuum.

There is a general agreement that the UV continuum of higher luminosity quasars is harder \citep[][however, see also \citealt{2008AJ....136.1799K}]{Carballo et al.(1999),Telfer et al.(2002),Davis et al.(2007)}. 
On the other hand, there has been a long running debate on the existence of the redshift dependence (evolution) of the quasar continuum.
Some studies suggest the slight hardening of the continuum toward higher redshift \citep{Carballo et al.(1999),2008AJ....136.1799K}, while others find hardly any evolution \citep{Kuhn et al.(2001),Pentericci et al.(2003)} or even reverse \citep{Wright(1981)}.

Several mechanisms could cause the redshift dependence of the quasar continuum, e.g., the evolution of the accretion process \citep{2008AJ....136.1799K} and/or the evolution of the dust component. 
Among them, one of the interesting mechanism is the intergalactic (cosmic) dust extinction, which makes the quasar spectra statistically redder at higher redshifts.
Indeed, \citet{Wright(1981)} attributed the redder spectra toward higher redshift they detected to the effect of the cosmic dust, and concluded that the average line of sight has cosmic extinction of $A_{\rm v} = 0.85\pm0.51$ out to z=3 (however, see also \citealt{Cheng et al.(1991)}). 
Compared to other attempts on probing the cosmic dust extinction \citep[e.g.,][]{More et al.(2009), 2010MNRAS.405.1025M}, quasars take advantage of their high luminosity and large sample size; therefore, they can be explored to higher redshifts.

In this Letter, we take a large quasar sample from Sloan Digital Sky Survey (SDSS, \citealt{York et al.(2000)}) and use the continuum slope of the quasar composite spectrum to probe its redshift dependence, then further explore its implications for the cosmic dust.

This Letter is organized as follows. In Section 2, we describe the data and method of building the composite spectrum. 
In Section 3, we report the finding of a significant dependence of $\alpha_\nu$ on the redshift: redder spectra at higher redshifts. 
In Section 4, we build a simple cosmic dust extinction model to quantify the reddening effects. 
In Section 5, we discuss other possibilities. Finally, we present our conclusions in Section 6.
Throughout this Letter, we adopt the cosmological parameters $\Omega_{\rm{\Lambda}}$ = 0.7, $\Omega_{\rm M}$= 0.3, and $h = 0.7$.

\section{Data}\label{sec:data}
The SDSS legacy survey provides a database of quasar samples that includes photometries in \textit{u,g,r,i}, and \textit{z} bands and spectra from 3800 to 9200 $\rm\AA$ with resolutions of about 2000 \citep{York et al.(2000)}. 
In the SDSS DR7 quasar catalog, \cite{Schneider et al.(2010)} compiled 105,783 quasars that are more luminous than $M_i=-22.0$ and have at least one emission line with FWHM larger than 1000 km s$^{-1}$ or have interesting/complex absorption features. 
This catalog provides the basic database for our study.

Accompanying the DR7 quasar catalog, there are other value-added catalogs. 
\cite{Hewett Wild(2010)} recalculated the redshift of quasars and reduced the systematic effects of the redshift to the level of 30 km s$^{-1}$ per unit redshift. 
\cite{Wild Hewett(2010)} improved the sky subtraction in the red wavelength band. After measuring the spectroscopic features, \cite{Shen et al.(2011)} calculated the implied physical parameters (e.g., the bolometric luminosity, black hole mass, and Eddington ratio) for each quasar.

In this study, we take the spectra from \cite{Wild Hewett(2010)} and use the improved redshifts from \cite{Hewett Wild(2010)}. 
We exclude the quasars with broad absorption lines from our analysis according to the flags in \cite{Shen et al.(2011)}. The radio-loud quasars show significant redder continua than radio-quiet ones \citep{Labita et al.(2008)} and the fraction of them changes with both redshift and luminosity \citep{Jiang et al.(2007)}. 
For these reasons, we also remove them.
Therefore, our final sample with calculated bolometric luminosity includes 91,131 objects.

\subsection{Composite spectrum}\label{sec:method}

We first correct the galactic reddening for each quasar using the SFD map \citep{Schlegel et al.(1998)} and the reddening curve of \citet{Cardelli et al.(1989)}. 
Then, we mask bad pixels and shift the spectra to their rest frame according to the redshifts of \cite{Hewett Wild(2010)}. 
Next, we follow \cite{Vanden Berk et al.(2001)} to make the quasar composite and calculate the statistical error for the composite spectrum by dividing 68$\%$ semi-quantile to the square of the contributing number at each wavelength.

\begin{figure}[ht!]
\hspace*{-1.cm}
\epsscale{1.3}
\plotone{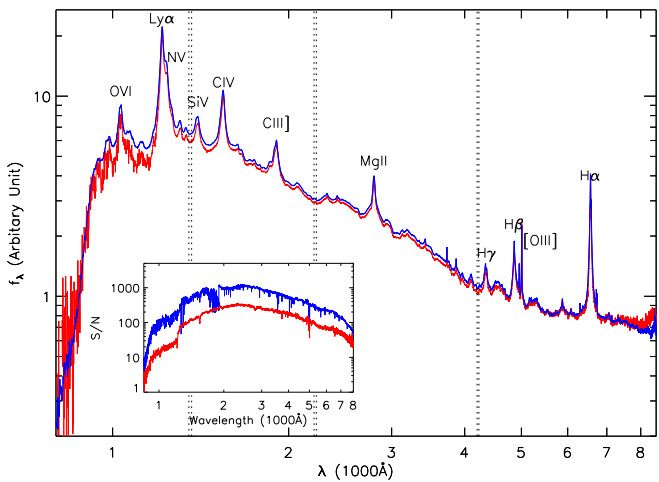}
\caption{
Composite quasar spectrum generated from 91,131 SDSS DR7 quasars (this work, blue line) and the result of \cite[][red line]{Vanden Berk et al.(2001)}.
The signal-to-noise ratios of these two composite quasar spectra are plotted in the sub-panel. 
The dotted lines mark the regions of continuum windows, i.e., 1350-1365 $\rm\AA$, 2210-2230$~\rm\AA$ and 4200-4230 $\rm\AA$, which are used to calculate $\alpha_{\nu12}$ and $\alpha_{\nu24}$ (see the text for details).}
\label{Fig1}
\end{figure}

The composite spectrum of all 91,131 quasars is shown in Fig.~\ref{Fig1} as the blue line. 
For comparison, the composite spectrum of \cite{Vanden Berk et al.(2001)} is shown as the red line.
As can be seen, these two composites show good consistency.
Because of a much larger sample size, our composite spectrum has a much higher signal-to-noise ratio (shown in the sub-panel of Fig.~\ref{Fig1}).

In the UV-Opt band, the quasar continuum can be very nicely characterized by a single power law ($f_{\nu}\propto\nu^{\alpha_\nu}$) and is usually fitted from continuum windows.
In \cite{Vanden Berk et al.(2001)}, two continuum windows, 1350-1365 $\rm\AA$ and 4200-4230 $\rm\AA$, have been adopted, which avoid the $\rm Ly\alpha$ line at shorter wavelengths and contamination from the host galaxy at longer wavelengths.
However, this range is too long to be covered by any single SDSS spectrum at any given redshift. 
Similar to \cite{Davis et al.(2007)}, we add a new continuum window 2210-2230 $\rm\AA$. 
With this configuration of continuum windows and the 3800-9200 $\rm\AA$ wavelength range of SDSS quasars, 
the far-UV slope $\alpha_{\nu12}$ and the near-UV slope $\alpha_{\nu24}$ can be fitted for quasars in the redshift intervals $1.80<z<3.15$ ($n=22251$) and $0.71<z<1.19$ ($n=18068$), respectively.

\section{Result}\label{sec:result}
\subsection{Trend of $\alpha_\nu$ with bolometric luminosity}\label{sec:L}
\begin{figure}[ht!]
\hspace*{-1.cm}
\epsscale{1.3}
\plotone{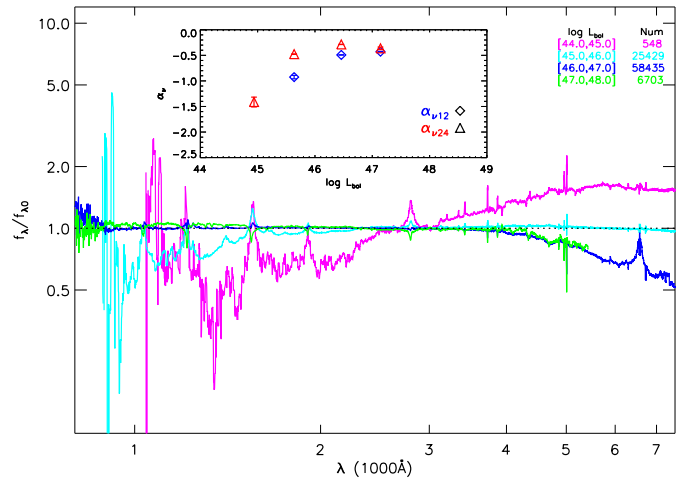}
\caption{Ratios of the composite spectra in different bolometric luminosity bins to that of the whole sample.
All the composites are normalized to the flux at 3000 $\rm \AA$.
The range of the bolometric luminosity bin (in $\rm{log}$ $L_{\rm{bol}}$) and the number of contributing spectra for each composite are listed in the legend.
In the sub-panel, the continuum slopes $\alpha_{\nu12}$ (diamonds) and $\alpha_{\nu24}$ (triangles) are plotted against the median bolometric luminosity of each bin.}
   \label{Fig2}
   \end{figure}

We bin the quasars into four bolometric luminosity ($L_{\rm{bol}}$) bins with a bin size of 1 dex from $10^{44}~\rm{erg~s}^{-1}$ to $10^{48}~\rm{erg~s}^{-1}$ and then make composites respectively. 
To show the possible variance of the composites more clearly, each composite spectrum is further divided by the global composite spectrum (blue line in Fig.~\ref{Fig1}) and normalized at 3000 $\rm\AA$. 
The results are shown in Fig.~\ref{Fig2}. 

The quasar composite spectra change with $L_{\rm{bol}}$ significantly, where the higher $L_{\rm{bol}}$ quasars have systematically harder continua. 
We calculate the continuum slopes ($\alpha_{\nu12}$ and $\alpha_{\nu24}$) for each composite spectrum and plot them as a function of $L_{\rm{bol}}$ in the sub-panel of the figure. 
The $\alpha_{\nu12}$ and $\alpha_{\nu24}$ values are close at high $L_{\rm{bol}}$ and show similar trends with luminosity. 
At the low luminosity side ($\rm{log}$ $L_{\rm{bol}}<47$), the change of the continuum slope with luminosity is very significant, while at the high luminosity side ($\rm{log}$ $L_{\rm{bol}}>47$), quasar continua roughly keep a constant slope ($-0.5$).

The redder spectra of lower luminosity quasars can hardly be explained by the accretion disk model, where the peak of the big blue bump moves toward lower frequency (longer wavelength) with increasing luminosity \citep{Hubeny et al.(2000), Davis et al.(2007)}. 
As discussed in \cite{Davis et al.(2007)}, this discrepancy could be alleviated by assuming a higher intrinsic dust extinction in low luminosity quasars. 
Indeed, there is observational evidence that the low luminosity AGNs have higher intrinsic dust extinction \citep[e.g.,][]{Gaskell et al.(2004)}. 
Such a luminosity dependent intrinsic dust extinction scenario is also consistent with a receding dust torus model \citep{Simpson(2005),Gu(2013)}. 
However, even if the dust extinction is independent of the quasar luminosity, quasars with higher extinction will also show lower luminosity and redder UV spectra. 
Such a selection effect may also partly explain the $\alpha_\nu -L_{\rm{bol}}$ relation we observed.

\subsection{Trend of $\alpha_\nu$ with redshift}
\begin{figure}[ht!]
\hspace*{-1.cm}
\epsscale{1.3}
\plotone{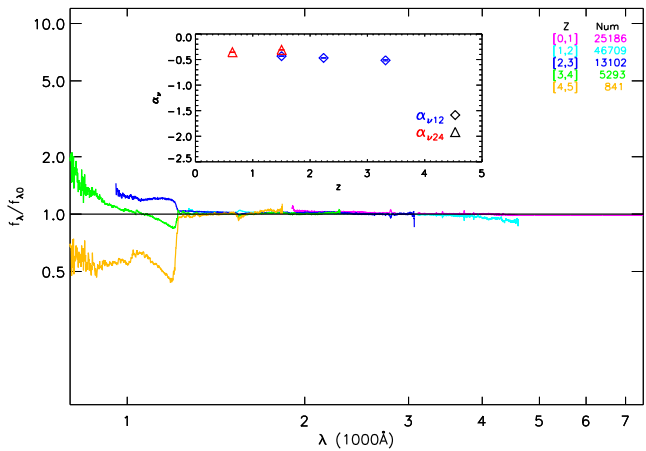}
\caption{Ratios of composite spectra from different redshift bins to that of the whole sample.}
   \label{Fig3}
   \end{figure}

In this section, we test whether the quasar continuum slope shows systematical variance with redshift. 
As a first step, we divide the sample into five redshift bins with bin sizes of 1 from $z=0$ to 5 and make composites. 
Similar to Fig. \ref{Fig2}, we calculate the ratios of the composites to the global one for different redshift bins and show the results in Fig. \ref{Fig3}. 
Consistent with the finding of \cite{Pentericci et al.(2003)}, we find little dependence of quasar spectra on redshift. 

However, the SDSS quasar sample is roughly a magnitude limit sample and thus high redshift quasars are biased to high luminosity ones.
According to the results in Section 3.1, if there are no evolutionary effects of quasar spectra, high redshift quasars should also have harder UV spectra. 
Therefore, the apparent non-evolutionary quasar spectra implies another redshift dependent mechanism that could compensate for the luminosity bias.
To recover this hidden effect, we make composite quasar spectra at different redshift bins after $L_{\rm{bol}}$ controlled.

\begin{figure}[ht!]
\hspace*{-1.cm}
\epsscale{1.3}
\plotone{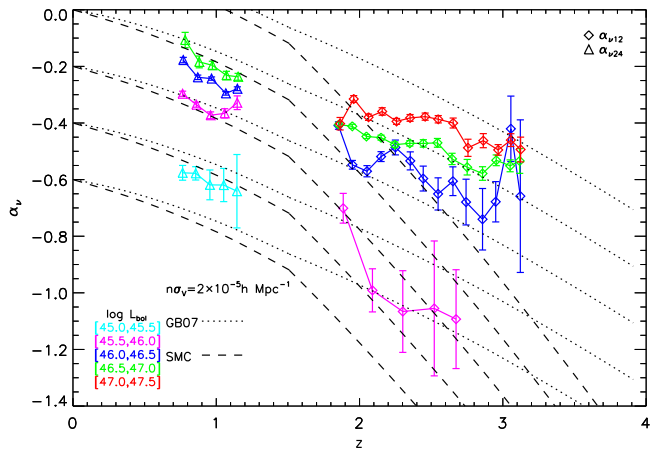}
\caption{$\alpha_{\nu12}$ (diamonds) and $\alpha_{\nu24}$ (triangles) of quasar composites in different $L_{\rm{bol}}$ bins (different colors) as a function of redshift.
Dust effective density $n\sigma_{\rm v}$ is fixed to $2\times10^{-5}h~\rm Mpc^{-1}$ and two extinction models: GB07 and SMC are drawn in dotted and dash lines.
$\alpha_{\nu}$ is probed by $\alpha_{\nu24}$ below $z=1.5$ and by $\alpha_{\nu12}$ above $z=1.5$.
}
\label{Fig4}
\end{figure}

The quasars are grouped into two-dimensional bins of $L_{\rm{bol}}$ and $z$.
The bin width of $L_{\rm{bol}}$ is set to 0.5 dex in logarithm while the width of $z$ is set to either 0.1 or 0.2 according to the number of the quasars available for composite. 
The minimum number of quasars to make a composite spectrum is set to 20. 
The results are shown in Fig.~\ref{Fig4}, where the spectral slopes are plotted against redshift with diamonds ($\alpha_{\nu12}$) and triangles ($\alpha_{\nu24}$).
The continuum slopes within the same $L_{\rm{bol}}$ bin are connected with lines, while different $L_{\rm{bol}}$ bins are represented by different colors.
We find that in any given $L_{\rm{bol}}$ bin, quasars at higher redshifts have systematically redder UV continuum slopes (for both $\alpha_{\nu12}$ and $\alpha_{\nu24}$).

This result clearly identifies the existence of the redshift dependence of quasar spectral slopes and implies that the previous debates are probably driven by redshift selection bias.

\section{Modeling the redshift dependence of the quasar continuum slope with cosmic dust extiction}\label{sec:model}
As we have mentioned in the Introduction, the cosmic dust extinction is a very interesting mechanism that can make the quasar continua systematically redder at high redshift. 
Therefore, it is quite natural to introduce a cosmic dust extinction model to explain the results we have obtained in Fig. ~\ref{Fig4}. Other possibilities will be discussed in Section 5.

We start from a toy model, which assumes that dust is uniformly distributed along the line of sight. 
Considering the intervening dust has a comoving number density of $n$, the optical depth of the cosmic dust to the photons with wavelength $\lambda_0$ emitted from a quasar at redshift $z_Q$ is
\begin{equation}\label{Equ_tau}
\tau(\lambda_0) = \int_0^{z_Q} n\sigma(\frac{1+z_Q}{1+z} \cdot \lambda_0)D_H\frac{(1+z)^2}{E(z)} dz
\end{equation}
where $D_H = c/H_0$, $E(z) = {H(z)}/{H_0}$, and $\sigma(\lambda)$ is the cross section of dust absorbers at wavelength $\lambda$ \citep{More et al.(2009)}.
We assume that the properties of the absorbers do not evolve with redshift and $\sigma(\lambda)$ is scaled by the cross section of the absorber at 5500$\rm\AA$, $\sigma_{\rm v}$:
\begin{equation}\label{Equ_sigma}
\sigma(\lambda)=\sigma_{\rm v} \cdot el(\lambda)/el(5500\rm\AA)\,
\end{equation} where el($\lambda$) stands for the extinction law.

We also assume that the intrinsic continuum of quasars in the UV-Opt band can be characterized by a pure power-law form $f_\nu=\nu^{\alpha_{\nu0}}$.
Quasars are put to their redshifts $z_Q$, and then, their observational spectral continuum can be rebuilt and slopes $\alpha_{\nu,z}$ can be measured after considering the cosmic dust extinction from $z=z_Q$ to $z=0$.
To mimic the observation, we also fit the continuum slopes in different wavelength ranges for quasars at different redshifts ($\alpha_{\nu12}$ for $z_Q>1.5$ and $\alpha_{\nu24}$ for $z_Q<1.5$).
In this toy model, besides the extinction law, we have two free parameters, one is the combined parameter $n\sigma_v$, which represents the effective density of the cosmic dust particles, and the other is $\alpha_{\nu0}$, the intrinsic quasar continuum slope.

We have little prior knowledge of the template extinction law of the cosmic dust. As the 2175 $\rm\AA$ extinction bump is rarely seen in other galaxies, an SMC-type extinction curve is preferred. 
In this study, we test the featureless SMC extinction law first. 
We choose sets of representative values of ${\alpha_{\nu0}}$ and show the predicted $\alpha_{\nu,z}$ as a function of $z_Q$, which are represented by the dashed parallel lines in Fig.~\ref{Fig4}. 
Here, $n\sigma_{\rm v}=2\times10^{-5}h~\rm Mpc^{-1}$ is tuned to match the observed $\alpha_{\nu24}-z$ relation.

As can be seen, our simple cosmic dust extinction model can reproduce the observed $\alpha_{\nu24}-z$ relation at the low redshift range $(z<1.5)$ very well. 
At high redshift, our model (SMC extinction curve and $n\sigma_{\rm v}=2\times10^{-5}h~\rm Mpc^{-1}$) predicts a much steeper $\alpha_{\nu12}-z$ relation than that is observed. 
This discrepancy can be easily solved by adopting either a lower $n\sigma_{\rm v}$ at $z>1.5$ or a shallower extinction curve at a far-UV wavelength. 
To keep our model simple, we test the later possibility. 
Indeed, there are studies suggesting that the extinction curves in quasars are flatter in the UV \citep{Czerny et al.(2004),Gaskell et al.(2004)}. 
We take the flat reddening curve from \cite[][hereafter, GB07]{2007arXiv0711.1013G}, which is quite similar to the Milky Way extinction law, except for the lack of a 2175 $\rm\AA$ bump.
We show the GB07 dust extinction models with the same $n\sigma_{\rm v}=2\times10^{-5}h~\rm Mpc^{-1}$ as the dotted lines in Fig.~\ref{Fig4}.
As expected, the GB07 extinction curve matches the observation much better than the SMC extinction curve at $z>1.5$.

\begin{figure}[ht!]
\hspace*{-1.cm}
\epsscale{1.3}
\plotone{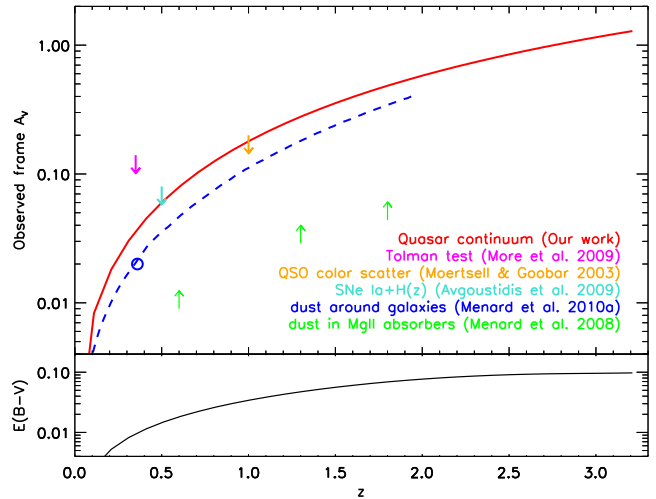}
\caption{Extinction magnitude $A_V$ as a function of redshift using the GB07 extinction model in Fig. ~\ref{Fig4}.
The open circle shows the result of \cite{2010MNRAS.405.1025M}, where the dashed line shows their extrapolation from a constant comoving dust density model. 
The other observational constraints are extracted from Fig. 9 of \cite{2010MNRAS.405.1025M}. 
In the bottom panel, the model-predicted dust reddening $E(B-V)$ is plotted against $z$.}
\label{Fig5}
\end{figure}

In order to have a more intuitive impression of the cosmic dust extinction, we take the GB07 dust extinction model and calculate the observer-frame extinction $A_V$ and reddening $E(B-V)$ as a function of source redshift.
The $A_V$ and $E(B-V)$ are plotted in the upper and lower panels of Fig.~\ref{Fig5} respectively. 
In this plot, we also show the cosmic extinction constrained from a few other measurements for comparison. 
The dashed line shows the constant comoving dust extinction model of \cite{2010MNRAS.405.1025M}, which is constrained from the statistical detection of dust reddening around galaxies up to large scales. 
The other observational constraints are also extracted from Fig. 9 of \cite{2010MNRAS.405.1025M}, which includes the results from measuring the excess scatter seen in higher redshift quasar colors 
\citep{2003JCAP...09..009M}, using the Tolman test \citep{More et al.(2009)}, combining constraints from luminosity distances and $\rm{H(z)}$ \citep{Avgoustidis et al.(2009)}, and 
using MgII absorbers \citep[][see Appendix C of \citealt{2010MNRAS.405.1025M} for details]{2008MNRAS.385.1053M}. 
Further study about the impact of cosmic dust on supernova cosmology can be found in \cite{Corasaniti(2006)} and \cite{2010MNRAS.406.1815M}.

\section{Discussion: other possibilities}
In this section, we discuss whether there are any other explanations to the redshift dependence of the quasar continuum slopes other than the cosmic dust.

\subsection{Evolution of intrinsic properties?}
In Section 3.1, we have argued that the internal dust extinction might be the cause of the $\alpha_\nu$-luminosity trend that we observed.
Could it also be responsible for the $\alpha_\nu -z $ relation?
In general, we know that metal and dust are accumulated through various processes in the galaxy evolution, e.g., dust formation in Type II supernova and asymptotic giant branch stars \citep{Asano et al.(2014)}.
Therefore, galaxies at lower redshifts will typically have more dust. 
This global trend is opposite to the $\alpha_\nu -z $ relation we observed for quasars.

However, we know that quasar host galaxies are atypical.
Many observations have shown the presence of large dust mass in high redshift $(z>4)$ quasars; though, the origin of the large amount of dust in such early epochs is unclear \citep[][and references therein]{Valiante et al.(2011)}.
Moreover, there are also clues that the extinction curves of redshift $(z>4)$ quasars might be different from their lower redshift counterparts \citep[e.g.,][]{Nozawa et al.(2015)}.
Despite the variations, we have not yet found any statistical evidence that quasars at higher redshifts show systematically higher intrinsic extinctions, at least for the redshift range we probed ($0.7<z<3.2$). 
Nevertheless, we emphasize that such a possibility still opens.

Besides the possible evolution of the intrinsic dust, could other physical properties (e.g., the black hole accretion process) of quasars evolve? 
Since the black hole accretion is a local physical process, we do not expect that they show significant redshift evolution effects once their basic physical parameters are fixed (e.g., black hole mass or accretion rate).
In this work, we group quasars into $L_{\rm{bol}}-z$ bins and do not further consider the variation of other physical parameters. 
However, we have tested that whether, even after constraining quasars to small black hole mass $M_{\rm{BH}}$ and Eddington ratio $L_{\rm{Edd}}/L_{\rm{bol}}$ ranges (0.1 dex) in the original 2D $L_{\rm{bol}}-z$ bins, the anti-correlation between $\alpha_\nu$ and redshift still exists.

\subsection{Host galaxy contamination?}
Quasars are hosted by galaxies, but they outshine them in most cases.
However, for the low luminosity quasars at low redshift, the contamination of starlight might still be significant, especially at the near-UV band, i.e., $\alpha_{\nu24}$.
For a $\rm{log}$ $L_{\rm{bol}}\sim45.75$ quasar, a bright star-forming host galaxy with $M_{\rm r}=-23$ contributes $\sim20$\% of the flux in the $2000\sim4000~\rm\AA$ range. 
More than that, the redshift range of the near-UV band we probed is from $z=0.7$ to $z=1.1$. 
This is the time when the cosmic star formation rate (SFR) starts to decrease. 
Therefore, it is possible that there are larger contributions of the starlight at higher redshift, so that makes the quasar spectra look redder.

To test this possibility, we generate a series of model galaxy spectra at different redshifts according to the cosmic star formation history given in \citet{2014ARA&A..52..415M} and their luminosities are assumed to be $M_{\rm r}=-23$. 
We then combine the galaxy spectra at different redshifts with the composite spectrum of $\rm{log}$ $L_{\rm{bol}}\sim 45.25$ quasars. 
Again, we measure the $\alpha_{\nu24}$ for the synthetic quasar+galaxy spectra at different redshifts. 
We find that, globally, $\alpha_{\nu24}$ becomes smaller (redder) when the galaxy spectra is combined into the quasar continuum. 
However, on the other hand, we do not find that the combined spectra get redder toward high redshift. 
The reason is that, while going to the higher redshifts from $z\sim 0.7$ to $1.1$, because of the higher SFR, not only  does the contribution of the starlight in the near-UV band become larger, 
but the near-UV band spectrum of the galaxy itself also becomes bluer.

\subsection{Quasar color selection bias?}
SDSS quasars are selected using a complicated color-based algorithm \citep{Richards et al.(2002)}.
Quasar candidates at high and low redshift are selected based on different color-color diagrams. 
Therefore, the color-selection criteria might introduce biases to the intrinsic colors of quasars at different redshifts.
For example, is it possible that the redder quasar continua at higher redshifts are caused by the missing of red quasars at low redshift or vice versa? 

To test this possibility, we take the reddest (the last magenta diamond in Fig.~\ref{Fig4}) composite spectrum and put it at different redshifts (from $z=3$ to 0) and calculate their observed colors. 
We do the same thing for the bluest one (the first green triangle). 
We find that the tracks of the colors of these two composites along the redshift still assemble in the quasar color-color selection area.
From this test, we argue that there is no significant selection bias in the colors of quasars at different redshifts in SDSS.

\section{Conclusion}
In this Letter, we use a large non-BAL radio-quiet quasar sample from SDSS DR7 to study the redshift and bolometric luminosity dependence of quasar continuum slope by means of the composite spectrum. 
Consistent with previous studies, we find that the UV-Opt spectrum becomes harder toward higher bolometric luminosity, and we suggest that the internal dust reddening might be the main reason for this luminosity dependence. 

When the bolometric luminosity is controlled, we find that the quasar UV continuum slope is anti-correlated with redshift (redder at higher redshift). 
We propose that the redshift dependence of the quasar UV slope might be caused by the reddening of the cosmic dust. 
We build a simple cosmic dust extinction model to quantify the reddening of the quasars. 
Based on our model, we find that an effective dust density $n\sigma_v \sim 10^{-5}h~\rm Mpc^{-1}$ at $z<1.5$ can reasonably reproduce the $\alpha_{\nu24}-z$ relation. 
The $\alpha_{\nu12}-z$ relation at high redshift ($z>1.5$) can also be easily reproduced by using a shallow extinction curve or adopting a smaller effective density of the cosmic dust. 
The cosmic dust extinction constrained from our model is globally consistent with the constrains from other independent measurements, and has reached to an unprecedented high redshift ($z\sim3.2$).

\acknowledgments
We thank the referee for a careful reading and highly appreciate the comments and suggestions.
X.X.Y. thanks Peng Jiang in USTC for help with our code writing, Hengxiao Guo and Fangting Yuan in SHAO for helpful discussions, and Brice M\'enard in JHU for providing their model data.
This work was supported by the Strategic Priority Research Program ``The Emergence of Cosmological Structures'' of the Chinese Academy of Sciences (CAS; grant XDB09030200),
the National Natural Science Foundation of China (NSFC) with the Project Numbers 11433003, 11390373, and 11103058, and the ``973 Program'' 2014 CB845705.


\end{document}